\def\year{2020}\relax
\documentclass[letterpaper]{article} % DO NOT CHANGE THIS
\usepackage{aaai20}  % DO NOT CHANGE THIS
\usepackage{times}  % DO NOT CHANGE THIS
\usepackage{helvet} % DO NOT CHANGE THIS
\usepackage{courier}  % DO NOT CHANGE THIS
\usepackage[hyphens]{url}  % DO NOT CHANGE THIS
\usepackage{graphicx} % DO NOT CHANGE THIS
\usepackage{amsmath}
\usepackage{graphicx}  % DO NOT CHANGE THIS
\author{Zhi Liu,Jing Gao, Li Chen\\
iLambda\\
Aurora, Ohio, USA \\
\{zliu, jgao, lchen\}@ilambda.ai,
\And
Yan Huang,\\
{University of North Texas}\\
Denton, Texas, USA \\
huangyan@unt.edu, 
\And
Dong Li\\
{Kent State University}\\
Kent, Ohio, USA \\
dli12@kent.edu}

\title{Large-scale Real-time Personalized Similar Product Recommendations }
\begin{document}

\maketitle

\begin{abstract}
Similar product recommendation is one of the most common scenes in e-commerce. Many recommendation algorithms such as item-to-item Collaborative Filtering are working on measuring item similarities. In this paper, we introduce our real-time personalized algorithm to model product similarity and real-time user interests. We also introduce several other baseline algorithms including an image-similarity-based method, item-to-item collaborative filtering, and item2vec, and compare them on our large-scale real-world e-commerce dataset. The algorithms which achieve good offline results are also tested on the online e-commerce website. Our personalized method achieves a 10\% improvement on the add-cart number in the real-world e-commerce scenario. 
\end{abstract}

\vspace{-3ex}
\section{Introduction}

The recommender system has been widely applied in e-commerce platforms. Traditional algorithms focus on generating personalized top-k item lists for users. However, in real-world platforms, the top-k recommendations are only applicable for part of scenarios. For example, in our e-commerce platform, the personalized top-k recommendation only can be used in the homepage recommendation, and only 4\% of user visitings come from the homepage. In this paper, we study the similar product recommendation, which contributes 33\% user visitings on our platform. Similar product recommendation is widely applied in different scenes on e-commerce platforms. When a user visits an item, a list of similar products will be presented to the user to help the user compare with other products and make a better choice, such as Amazon's related item recommendation and eBay's similar sponsored items.

Existing methods such as content-based similarity, item-to-item collaborative filtering, and some user-based approach can be used in similar product recommendations. However, such algorithms are designed for similarity calculation and applied in general recommendation scenes such as personalized recommendations based on user activities. Without considering the recommendation scenario, these algorithms miss one or more important points: the similar product lists are non-personalized, cannot process user real-time activities and generate the most recent interests, inflexible when dealing with special requirements, and item similarity and user interests are not combined. In our model, many key factors in real-world recommender systems such as real-time user interests, balancing the influence of user interests and item similarity, time efficiency, and other business requirements, will be considered instead of calculating the similarities between items only.

Different from personalized recommendation, similar products calculation is focused on the current item instead of user interests. User behaviors will be only considered in the calculation of similarities between items in most algorithms\cite{linden2003amazon}. In some ways, it is reasonable since similar product recommendations highly concentrate on the current interests of users and this is closely related to the item the users are visiting. Additionally, a non-personalized recommendation can also help users explore different and new items they have not seen before instead of falling into a small set of items generated by personalized recommendation algorithms. However, personalized recommendation algorithms can significantly improve the user experience by providing the items which are best matched with user interests. From our dataset, there are totally 33\% of user clicks come from non-personalized similar product recommendations, but only 25\% add-cart events are generated from these clicks. Compared with our personalized recommendation on the homepage (4\% click generates 6\% add-cart), the converting ratio from click to add-cart is much lower. So a personalized method in similar product recommendation is necessary for our system. Since such a recommendation scenario is highly affected by user current interests, a personalized model also needs to process real-time user activities. To design a personalized algorithm here, we have the following challenges:

\begin{itemize}
\item Both current visiting items and historical behaviors of users need to be considered. The algorithm must balance the influence of the current item and user interests.
\item The algorithm needs to deal with real-time user behaviors. Since similar product recommendations are highly affected by short term user interests, we need to use the most recent user visiting history to generate user interests vector.
\item The recommendation results can show more items to users in addition to the items generated by traditional personalized algorithms. Users can explore more items and will not be limited to a small set of items.
\item Similay product recommendation is one of our busiest channels. Each of our online servers needs to respond to several thousand requests per second with the response time less than 100 ms. So the efficiency of the real-time personalized model must be considered.
\item The model needs to meet the business requirements such as cross-category recommendation, and new item recommendation.
\end{itemize}

To solve the challenges above, we propose a framework with two components. In the first step, we combine the results from different offline algorithms and business requirements to generate a similar product pool for each item. Then we build an online real-time algorithm to provide the ranking results based on the current visiting item and user short-term interests. In this paper, we will briefly introduce our method of the first step, and then we focus on building and analyzing different ranking algorithms. 

We first introduce the most related research work. Then we will propose three baseline algorithms and our online personalized algorithm. Finally, we compare these algorithms on our real-world e-commerce dataset and show the online performance of these algorithms. 

\vspace{-2ex}
\section{Related Work}

The item-based collaborative filtering \cite{sarwar2001item} has been applied in the recommender system for decades. User interests and item similarity can be calculated by the user preference on different items. In a similar production calculation, one of the most important similarity product recommendation algorithm comes from Amazon \cite{linden2003amazon} and \cite{smith2017two}. The authors discuss the traditional collaborative filtering, cluster and search-based models, and then provide the algorithm calculating the item similarities from the user purchasing records. Inspired by the word2vec algorithm \cite{mikolov2013distributed}, \cite{barkan2016item2vec} proposes the item2vec algorithm. Items are represented by embeddings. Based on user activities, vectors can be trained and the inner product can be used to measure the similarity between items. In \cite{kabbur2013fism}, the authors propose a user-based CF algorithm by modeling the user ranking with the inner product of user and item vectors. In this paper, we apply the item-to-item CF and item2vec algorithms as our baseline methods and add some changes to fit our dataset.

In \cite{brovman2016optimizing}, the authors divide the recommendation into two steps, recall and ranking, and use the comparison features and item quality features in the model. Based on the assumption that historical items have different contributions to the current visiting item,  \cite{he2018nais} propose a neural network method for item-to-item collaborative filtering. In \cite{agarwal2018personalizing}, the authors propose a similar idea to our work. After generating a set of similar products, an ALS-based method and a Bayesian Personalized Ranking method will be applied to provide personalized ranking results. However, the personalized method is an offline method without real-time user behaviors. In this paper, we also apply the same framework with \cite{brovman2016optimizing} and focus on the ranking step. The user interests will be generated by most recent user activities.

Deep neural networks like RNN/LSTM, which capture the inherent sequential structures of data, have achieved promising successes in the natural language process field. It is flexible and natural to applying such sequential neural networks on recommendation tasks to mining the temporal dynamic features of user behaviors. \cite{hidasi2015session} proposes a session-based recommendation model. Taking the one-hot sequential encoding as input, the model outputs the likelihood for each possible item in the session. Unlike the previous session-based model without learning user representation, \cite{wu2017recurrent} proposes a RNN-based model,  which is capable of imitating the changes of the user interests and item features over time.

Neural network personalized recommendation is widely used in recent years. In \cite{covington2016deep}, the authors demonstrate the YouTube recommender system. Videos and different features are embedded and used as the input of the multilayer perceptron. Based on the attention model, \cite{zhou2018deep} proposes the neural network model to handle different features and the closeness between items. The wide and deep framework \cite{cheng2016wide} are also introduced in many personalized recommendation algorithms. In this paper, we build a neural network model to generate user current interests in real-time. Then based on the candidate pool, the model can provide a personalized similar product recommendation.

\vspace{-2ex}
\section{Baseline Algorithms}
%贡献，针对这个场景，实时，快，真在线算法，其他算法不反应实时用户选择

In this section, we propose three baseline algorithms include an item-to-item collaborative filtering algorithm, an image-similarity-based algorithm, and the item2vec algorithm. Limited by our online business rules, we will first generate a candidate pool for similar product recommendations. The items in the candidate pool come from there three different sources: product with similar attributes, similar new products, and products generated by some algorithms. Each item has a pool with 200 similar products. In the ranking step, we grade the products in the pool and show the top 30 items to customers. We skip the details of generating the pool and focus on the ranking algorithms.  

\vspace{-2ex}
\subsection{Image-Similarity-Based Algorithm}

We first introduce an imaged-similarity-based model. Clothes take the largest part of our e-commerce dataset. For the fashion products, the style, color, and lots of features which can be directly extracted from the picture play an important role when user making selections. So in our first model, we try to use the similarity between item image to generate recommendation results. Here we use the transfer learning method to generate a feature vector for each product image. Transfer learning is an effective way to extract features from images by a pre-trained model. In the experiment, we implement the algorithm by the ResNet152 pre-trained model from TensorFlow Keras. We remove the last dense layer and apply average pooling to get a vector with the length of 2048 for each product image. Then we rank the similarity between items by the value of cosine similarity.

\vspace{-2ex}
\subsection{Item-to-Item Collaborative Filtering}
Inspired by the Amazon item-to-item CF, we implement our baseline algorithm on the user behavior dataset. Different from previous work, we use the user clicks data to generate similar products since the order and rating data take a very small portion of the whole user behavior dataset (less than 1\%). Here we define the similarity between two items as:

\vspace{-1.7ex}
\begin{equation}
C_{sim}(v_i, v_j) = \frac{|U_{v_i} \cap U_{v_j}|}{|U_{v_i} \cup U_{v_j}|}
\end{equation}

Here the $U_{v_i}$ means the set of users who have visited item $v_i$. To enhence the variety of the recommendation results and deal with the cold start problem, we also add a feature-based cosine similarity as: 
\vspace{-2ex}
\begin{equation}
F_{sim}(v_i, v_j) = \frac{|F_{v_i} \cap F_{v_j}|}{|F_{v_i} \cup F_{v_j}|}
\end{equation}

The $F_{v_i}$ is the set of attributes of item $v_i$. Only the attributes which are shown to users will be used at here. Then we combine these two similarity together:

\vspace{-2ex}
\begin{equation}
sim(v_i, v_j) = \alpha C_{sim} + (1-\alpha) F_{sim}
\end{equation}

In our experiment, we set $\alpha$ as 0.9.

\vspace{-2ex}
\subsection{Item2vec}

In the skip-gram model, each word is represented as a vector. Following the same idea, \cite{barkan2016item2vec} provides the algorithm on similar product recommendations. In this paper, we build the model based on our user activities data. 
\vspace{-3ex}
\begin{figure}[ht!]
        \centering
        \includegraphics[height=2.4cm,trim=1.2cm .8cm .8cm .5cm,clip]{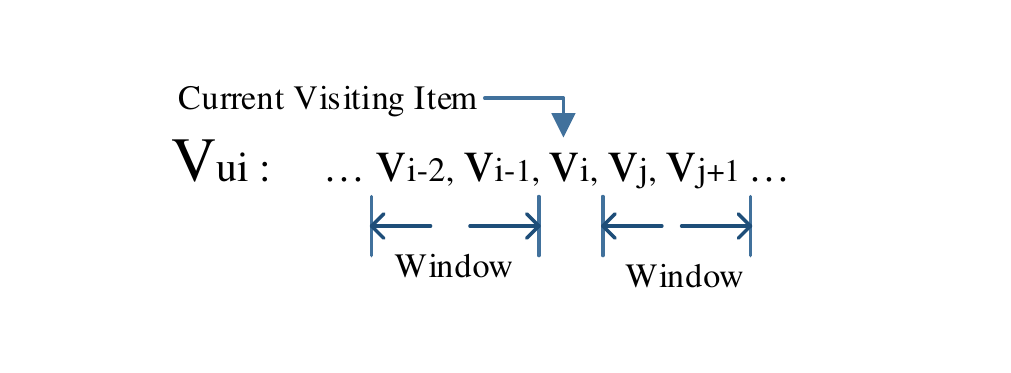}  
        \vspace{-2ex}
        \caption{Item2vec model}
        \label{fig:item2vec}
\end{figure}
\vspace{-2ex}
Given a set of items $V = \{v_i\}_{i=1}^{|W|}$, for each user $u_i$ from the user set $U$, we record the visiting behaviors of the user as $V_{u_i} = \{v_m, v_{m+1}, ... , v_n\}$ (sorted by visiting time). Then we choose the time window $l$, for $v_i \in V_{u_i}$, if the distance between the position of $v_i$ and another $v_j \in V_{u_i}$ is less than $l$, we say that $v_j$ in the window of $v_i$. The set of such $v_j$s is noted as $L_{v_i}$. 

The vector of item $v_i$ is noted as $\overrightarrow{v_i}$. For each item $v_i$, we randomly sample $n$ items from $V$ and define the loss function by sigmoid cross entropy: 

\begin{equation}
p(v_j|v_i) = \sigma(\overrightarrow{v_i}^T \overrightarrow{v_j} + b_j) 
\end{equation}
\vspace{-2ex}
\begin{equation}
\label{eq:loss}
L_{v_i} = -log(p(v_j|v_i)) - \sum_{v_k \in V_{neg}}log(1 - p(v_k|v_i))
\end{equation}

The $\sigma$ represents the sigmoid function, $\sigma(x) = 1/(1 + exp(-x))$, $v_j$ is the item in the window of $v_i$, and $V_{neg}$ is a set of negative sampling items for $v_i$. $b_j$ is the bias value of item $v_j$. In our experiment, we set the length of the time window with 2, and the number of negative sampling for each positive case is 8. When training the model, we scan every click in $V_{u_i}$. For each $v_i$ in $V_{u_i}$, the previous and following two items $v_j$ will be selected as a positive case. For each case, we randomly select 8 negative items from $V$ and apply the equation \ref{eq:loss} as the loss function.

In this model, we need to notice that actually we have two different items embeddings. The vector of the current visiting item comes from the first embedding, and the vectors of items in the window and negative sampling set come from the second embedding, here we call it weight embedding. The second embedding is only used in training. When grading the candidates, we use the cosine similarities between vectors in the first embedding.

\vspace{-2ex}
\section{Personalized Similar Product Recommendation}

In this section we introduce our personalized recommendation algorithm and the add-cart enhance method. In personalized recommendation, there are some common ways to generate the user vectors such as representing user interests by a vector and train the user vector, or pooling and concatenating from a neural network with user activities as the input. In our model, we generate the user vector by the weighted average pooling from the vectors of the most recent visited items. 
\vspace{-2ex}
\begin{figure}[ht!]
        \centering
        \includegraphics[height=4.6cm,trim=.8cm .8cm .5cm .5cm,clip]{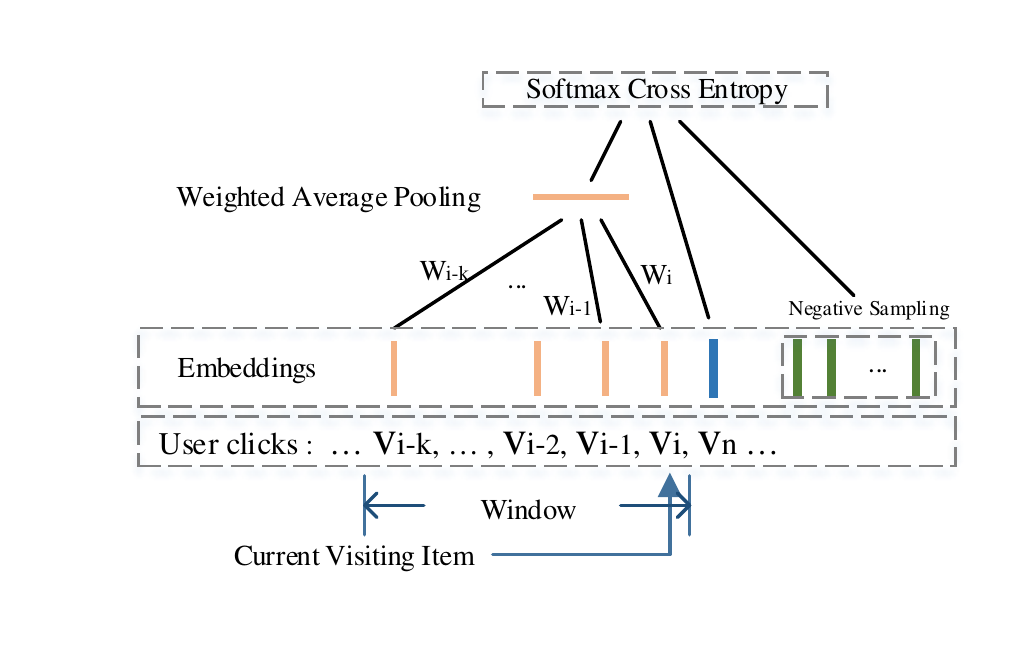}  
        \vspace{-2ex}
        \caption{Personalized similar product recommendation}
        \label{fig:model}
\end{figure}
\vspace{-3ex}

In the training process, we scan each item in $V_{u_i}$ for each user $u_i$. As it shown in figure \ref{fig:model}, for a $v_i \in V_{u_i}$, we set of a window $W_{v_i} = \{v_{j}, ..., v_i\}$ includes the current visiting item $v_i$ and $n$ items visited in previous of $v_i$. In our algorithm, we set the window size as 8. Since items in different positions have different influences on the current user interests, we assign a weight $w$ to each item in the set $W$, and the values of weights will be trained together with the embeddings. The current user vector can be calculated as:
\vspace{-1.2ex}
\begin{equation}
\label{eq:uvector}
\overrightarrow{u_i} = \frac{1}{n} \sum_{v_j \in W_{v_i}} w_j \overrightarrow{v_j}
\end{equation}
\vspace{-2ex}

The current user interests on item $v_i$ are modeled as:
\vspace{-0.7ex}
\begin{equation}
\label{eq:up}
p(v_i|u_i) = \overrightarrow{u_i}^T \overrightarrow{v_i}
\end{equation}
\vspace{-3ex}

Then we apply the current user interests in the personalized model. The loss function is defined by the softmax cross-entropy in forluma \ref{eq:loss2} with $v_n$ is the next item the user visit after $v_i$ and $V_{neg}$ is a set of negative sampling items as figure \ref{fig:model}.
\vspace{-2ex}
\begin{equation}
\label{eq:loss2}
L_{u_i,v_i} = -log(\frac{exp(\overrightarrow{u_i}^T \overrightarrow{v_n})}{exp(\overrightarrow{u_i}^T \overrightarrow{v_n}) + \sum_{v_k \in V_{neg}}exp(\overrightarrow{u_i}^T \overrightarrow{v_k}) })
\end{equation}
\vspace{-2ex}

In this model, we also have two item embeddings. When generating the vector $\overrightarrow{u_i}$, we look up the vectors of the items in the window from the first embedding. The vectors of items in $V_{neg}$ and the next item $v_n$ are from the other embedding. When grading the score, we will follow the formula \ref{eq:uvector} and \ref{eq:up}.

Another important user activity is the add-cart event. It is a stronger signal shows the user interests than common visiting, and generating more add-cart events is also closer to the final target of the recommender systems, improving the order number. Based on the algorithm we discuss above, we enhance the influence of the add-cart event in our model by defining the loss function as formula \ref{eq:loss3}:
\vspace{-2ex}
\begin{equation}
\label{eq:loss3}
AL_{u_i,v_i} = \omega L_{u_i,v_i}
\end{equation}

The parameter of $\omega$ is used to enhance the add-cart events. In the training process, we collect the items which the user finally add them to the cart. When the next item $v_n$ is in the add-cart set, we assign $\omega$ a larger value (2 in our experiment), if not, we set $\omega$ as 1.

\vspace{-2ex}
\section{Experiment}

In our experiment, we collect real-world e-commerce data to test different models. In the dataset, there are totally of 200 million user clicks come from 2 million users. 700,000 items are visited by these users. All of the data are generated in 8 days. In the experiment, we use the data from the first 7 days as our training data and test the models on the data from the last day. 

The item-to-item CF algorithm is implemented by Spark. To handle the large number of user visiting, we process different users parallel on different servers and then reduce the co-visiting items pairs for every item to generate the similarity score. The other three algorithms, image-based model, item2vec, and our personalized model are implemented by TensorFlow 2.0. 

\vspace{-2ex}
\subsection{Top K Hit Ratio}
The first experiment is calculating the top K hit ratio. For each click in the testing data, we get the size 200 candidate pool of the item and then apply our models to rank the 200 items. Then we record whether the next item visited by the user is contained by the top K items. The top K hit ratio is the total hit number divided by the total number of testing cases $T$ as formula \ref{eq:mea} and \ref{eq:mea2}.

\vspace{-2ex}
\begin{equation}
\label{eq:mea}
h(v_i) = 
\left\{
\begin{aligned}
1 & , & v_i \in topK \ list \\
0 & , & else 
\end{aligned}
\right.
\end{equation}

\vspace{-1ex}
\begin{equation}
\label{eq:mea2}
Acc = \frac{\sum_{v_i \in T} h(v_i)}{|T|}
\end{equation}
\vspace{-0.5ex}

\begin{table}[ht!]
\centering
\begin{tabular}{|c|c|c|c|}  
\hline
Methods &  Top 5 & Top 10 & Top 20 \\
\hline
Image-Similarity &  10.3 & 17.0 & 25.9 \\
\hline
Item-to-item  & 40.9 & 50.7 & 61.8 \\
\hline
Item2vec  & 42.4 & 53.3 & 61.0 \\
\hline
Personalized  & 45.6 & 55.2 & 64.1 \\
\hline
\end{tabular}
\vspace{-2ex}
\caption{TopK hit ratio (\%).}
\label{tab:exp}
\end{table}

In the experiment, we compare the top 5, 10, and 20 accuracies between different algorithms. The table \ref{tab:exp} shows the results of the click hit ratio. The personalized method achieves the best performance. Compared with item-to-item and item2vec, the personalized method improves the hit ratio by about 3 to 5\%.

\begin{table}[ht!]
\centering
\begin{tabular}{|c|c|c|c|}  
\hline
Methods &  Top 5 & Top 10 & Top 20 \\
\hline
Image-Similarity &  12.3 & 16.5 & 22.2 \\
\hline
Item-to-item  & 41.2 & 47.6 & 53.6 \\
\hline
Item2vec  & 39.7 & 43.4 & 52.0 \\
\hline
Personalized  & 47.2 & 54.4 & 66.1 \\
\hline
Add-Cart Enhance  & 53.7 & 61.6 & 70.2 \\
\hline
\end{tabular}
\vspace{-2ex}
\caption{TopK add-cart hit ratio (\%).}
\label{tab:exp2}
\vspace{-2ex}
\end{table}

Then we test the add-cart hit ratio of these algorithms. In this experiment, the settings are kept the same as the first experiment. The only difference is only the items which are added to the cart by users will be counted in the hit ratio. The table \ref{tab:exp2} shows the results. When we enhance the influence of the add-cart event, the hit ratio is significantly improved. The baseline algorithms perform worse than their click hit ratio, and the base personalized method has similar performance on add cart and click prediction. 

\vspace{-2ex}
\subsection{Online Performance}
\vspace{-1ex}

Finally, we apply the item-to-item CF model, item2vec model, and add-cart enhance model on the real-world e-commerce platform. The results show that the item-to-item CF model and item2vec model achieve similar online performance.  The improvement of the add-cart enhance model is less than the performance on the offline testing, but it still improves the online add-cart number by about 10\%. The model can also meet the efficiency requirement. We deploy the model by the TensorFlow estimator and use CPU only. Each server can respond to 2,000 requests per second with 16 cores and 128G memory. 
\vspace{-2ex}

\begin{figure}[ht!]
	\centering
	\includegraphics[ width=1\linewidth]{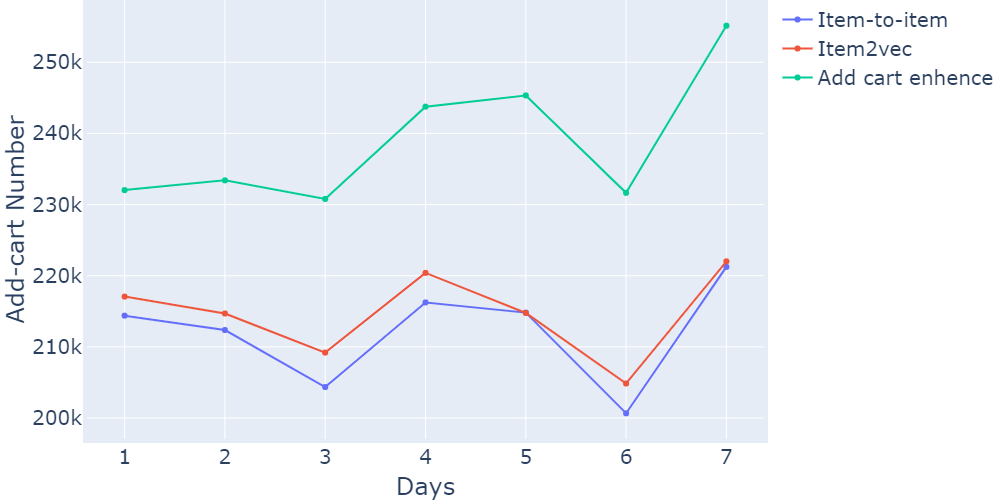}
	\vspace{-3ex}
	\label{fig:udis}

	\caption{Online testing results.}
	%\label{fig:exp}
\vspace{-2ex}
\end{figure}
\vspace{-1ex}

\vspace{-2ex}
\section{Conclusion}

In this paper, we introduce our personalized similar product recommendation algorithm and compare our method with some related work. The model will process real-time user activities, generate the most recent user interests, and ranking the similar items in the given candidate pool. Three models are tested on a real-world e-commerce platform, and the results show our model can significantly improve the performance of the add-cart number. In the future, we will try to model the influence of more factors in the algorithm like lacking information of new products, attributes, and the price.

\bibliographystyle{aaai}
\bibliography{citation}

\end{document}